\input{aipcheck}

\documentclass[
  ]
  {aipproc}

\layoutstyle{6x9}

\begin{document}

\title{Associated photon and heavy quark production \\
at high energy within $k_T$-factorization}


\author{A.V. Lipatov}{
  address={SINP, Moscow State University, 119991 Moscow, Russia}
}

\author{M.A. Malyshev}{
address={SINP, Moscow State University, 119991 Moscow, Russia}
}

\author{\underline{N.P. Zotov}}{
  address={SINP, Moscow State University, 119991 Moscow, Russia}
}

\begin{abstract}
 In the framework of the $k_T$-factorization approach, the production of prompt photons in association with a heavy (charm or beauty) quarks at high energies is studied. The consideration is based on the $O(\alpha \alpha_s^2)$ off-shell amplitudes of gluon-gluon fusion and quark-(anti)quark interaction subprocesses. The unintegrated parton densities in a proton are determined using the Kimber-Martin-Ryskin prescription. Our numerical predictions are compared with the D0 and CDF experimental data. Also we extend our results to LHC energies.

\end{abstract}
\maketitle


\section{Introduction}
Recently the D0 and CDF Collaborations reported data
\cite{D0,CDF,D0c} on associated (with a heavy quark jets)
prompt photon production at the Tevatron. 
The D0 Collaboration showed that the measured cross sections are in agreement with the
NLO QCD predictions~\cite{QCD} within theoretical and experimental
uncertainties in the region up to $p_T^{\gamma} \sim 70$ GeV. However, the substantial disagreement between theory and data for both
$\gamma + b$-jet and $\gamma + c$-jet production at large $p_T^\gamma$ was observed.
 The cross section slopes in data significantly differ from the predicted ones. The results indicate
a need for higher order perturbative QCD corrections
in the large $p_T^{\gamma}$ region.

 In the D0 papers~\cite{D0,D0c} it was demonstrated also that 
the $k_T$-factorization predictions~\cite{LMZ}
 are in a better agreement with the data.
   
 First application of $k_T$-factorization approach to production of photons associated with the charm
or beauty quarks have been performed in our previous paper~\cite{BLZ}. The consideration
was based on the ${\cal O}(\alpha \alpha_s^2)$ amplitude for the production of a single photon associated with a
quark pair in the fusion of two off-shell gluons $g^*g^* \to \gamma Q\bar Q$.
A good agreement between the
numerical predictions and the Tevatron data was obtained in the region of relatively low $p_T^\gamma$ where off-shell gluon fusion dominates.
However, the quark-induced subprocesses become more important at moderate and large $p_T^\gamma$
and therefore should be taken into account.
 Here we extend a previous predictions~\cite{BLZ} by including into
the consideration two additional ${\cal O}(\alpha \alpha_s^2)$ subprocesses, namely $q \bar q \to \gamma Q \bar Q$ and
$q(\bar q)Q \to \gamma q(\bar q)Q$, where $Q$ is the charm or beauty quark~\cite{LMZ}.

\section{Theoretical framework}
 According to the $k_T$-factorization theorem,
the cross section of the prompt photon and associated heavy quark production  can be
written as a convolution of the relevant off-shell
partonic cross sections and unintegrated parton
distribution functions (uPDF) in the proton $f_{i,j}(x,{\mathbf k}_{T}^2,\mu^2)$:
$$
  \sigma = \sum_{i,j} \int {\hat \sigma}_{ij}(x_1, x_2, {\mathbf k}_{1
T}^2, {\mathbf k}_{2T}^2) \, f_i(x_1,{\mathbf k}_{1T}^2,\mu^2) f_j(x_2,{\mathbf
k}_{2T}^2,\mu^2) \, dx_1 dx_2 \, d{\mathbf k}_{1T}^2 d{\mathbf k}_{2T}^2\frac{d\phi_1}{2\pi}\frac{d\phi_2}{2\pi},
$$

\noindent
where ${\hat \sigma}_{ij}(x_1, x_2, {\mathbf k}_{1T}^2, {\mathbf k}_{2T}^2), (i,j = q, g)$
is the relevant partonic cross section. The initial off-shell partons have fractions $x_1$ and $x_2$ of
initial protons longitudinal momenta, non-zero transverse momenta ${\mathbf k}_{1T}$ and ${\mathbf k}_{2T}$ and azimuthal angles $\phi_1$ and $\phi_2$.

In what concerns the uPDF, we took them in the KMR form~\cite{KMR}. The KMR formalism is a prescription for constructing the uPDF from the known standard PDF \footnote{Numerically, we used the MSTW-2008 set~\cite{MSTW} in the proton as the input.}. It gives $k_T$-dependent uPDF for both gluon and quark.

  The analytic expressions of the corresponding off-shell matrix elements were listed in~\cite{LMZ}. In the $k_T$-factorization approach the gluon polarization density matrix takes so called BFKL form: $\sum \epsilon^\mu\epsilon^{*\nu}=k^\mu_T k^\nu_T/\mathbf k^2_T$. The spin density matrix for the off-shell spinors is taken in the form $u(q)\bar u(q) = x\hat p$, where $q$ and $p$ are the quark and the proton momenta in the small $x$ and massless approximation~\cite{LMZ}.

 \begin{figure}[!b]
  \includegraphics[height=.27\textheight]{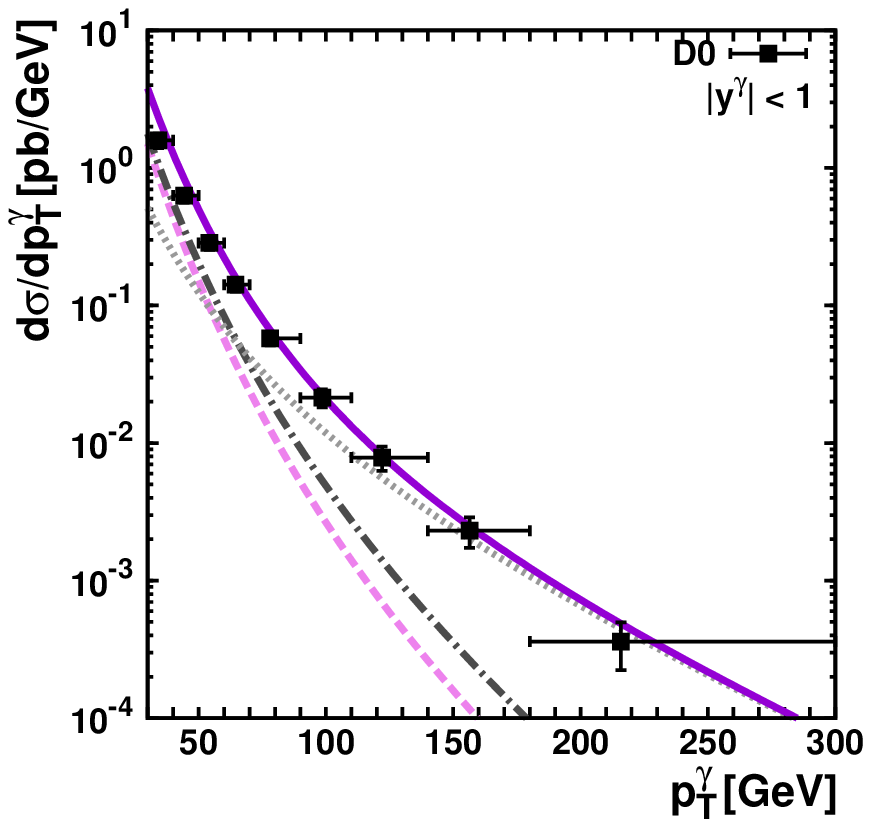}
  \includegraphics[height=.27\textheight]{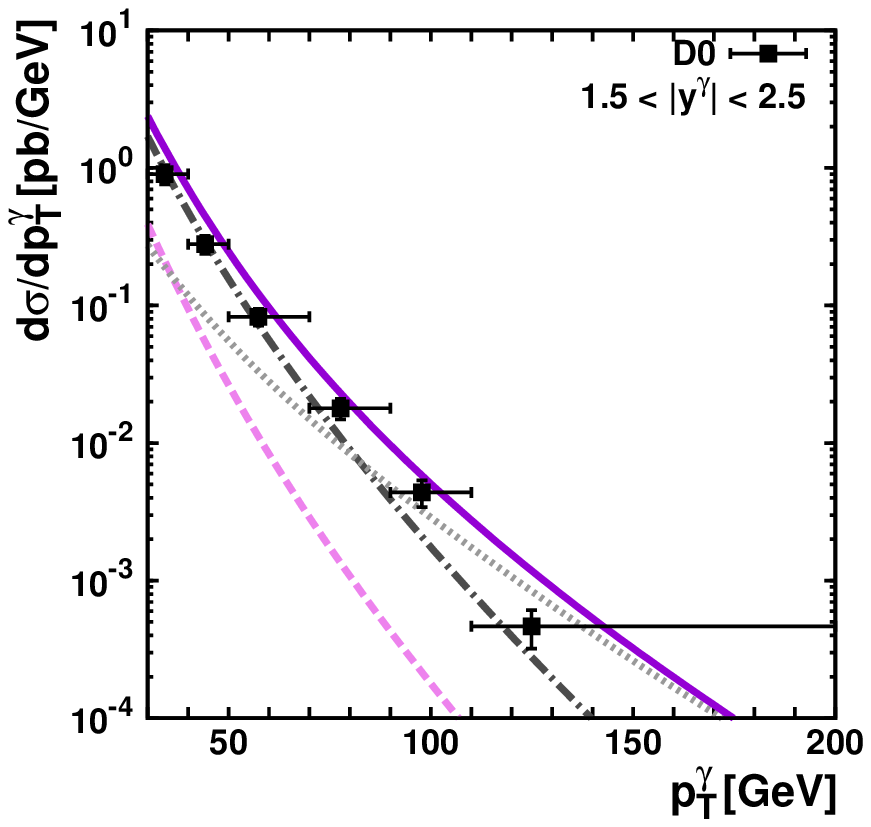}
  \caption{Differential cross section $d\sigma/dp_T^{\gamma}$ of associated $\gamma + b-jet$ production at $\sqrt s = 1960$ GeV, $|y^{jet}| < 1.5$ and $p_T^{jet} > 15$ GeV. The dashed, dotted and dash-dotted curves correspond to the contributions of $gg  \to \gamma Q \bar Q, q\bar q \to \gamma Q\bar Q, q(\bar q) Q \to \gamma q(\bar q)Q$ subprocesses. The solid curve represents their sum. The experimental data are from 
~\cite{D0}.
  }
  \label{fig1}
\end{figure}
In our numerical calculations we took the renormalization and factorization scales $\mu_R^2=\mu_F^2=\xi^2p_T^2$. In order to evaluate theoretical uncertainties, we varied $\xi$ between 1/2 and 2 about the default value $\xi=1$.
We used the LO formula for the strong coupling constant $\alpha_s(\mu^2)$ with $n_f=4$ active quark flavours at $\Lambda_{QCD}=200$ MeV, so that $\alpha_S(M_Z) = 0.1232
$. We set the charm and beauty quark masses to $m_c=1.5$ GeV and $m_b=4.75$ GeV.
\begin{figure}[!b]
  \includegraphics[height=.27\textheight]{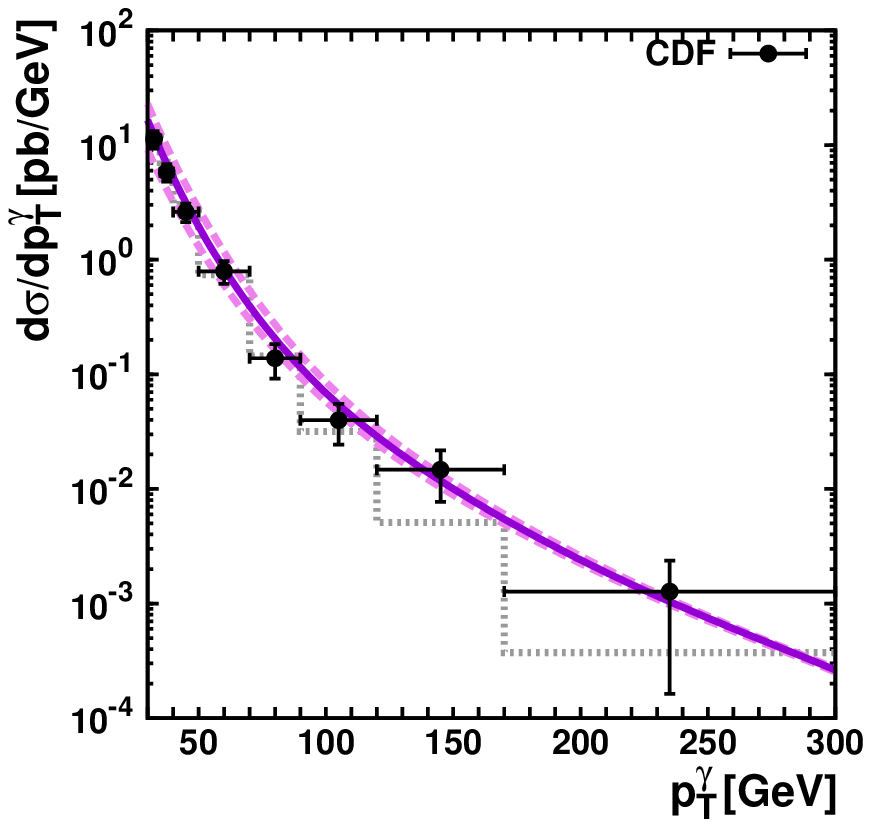}
  \includegraphics[height=.27\textheight]{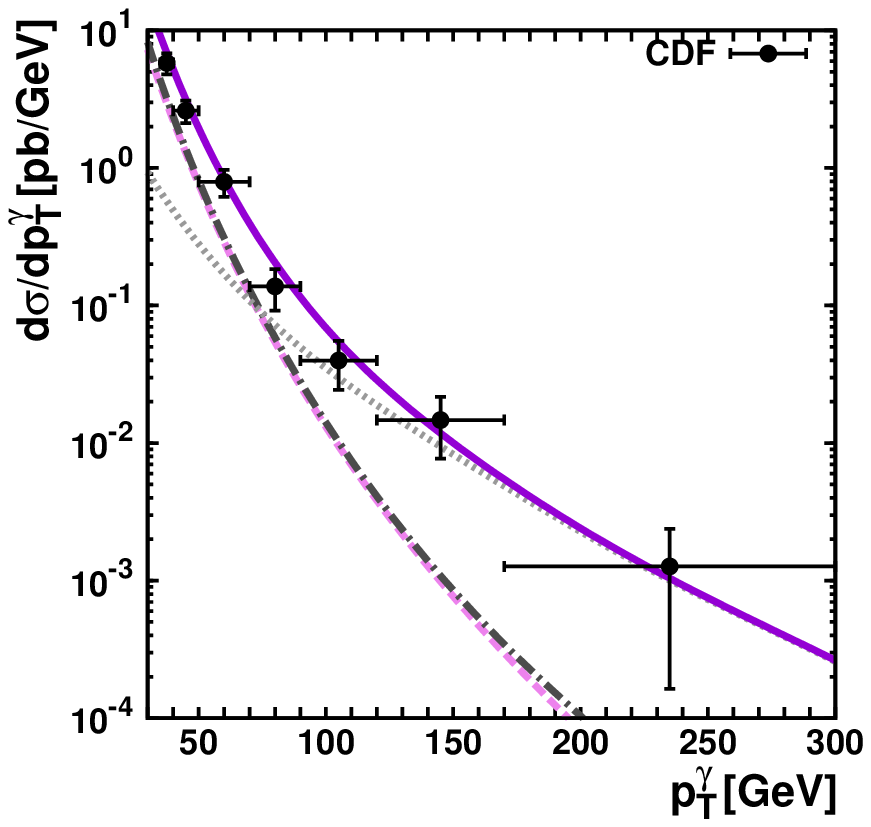}
  \caption{Differential cross section $d\sigma/dp_T^{\gamma}$ of associated $\gamma + c-jet$ production at $\sqrt s = 1960$ GeV, $y^{\gamma} <1, |y^{jet}| < 1.5$ and $p_T^{jet} > 20$ GeV. Notaion of all curves on right panel is the same as in Fig. 1.
The dotted histogram is the NLO pQCD predictions~\cite{QCD} taken from~\cite{CDF}. The experimental data are from ~\cite{CDF}.
  }
  \label{fig2}
\end{figure}
We use the experimental isolation cut for produced photons~\cite{D0,CDF,D0c}: 
$$
E_T^{had}\le E^{max}
$$
$$
(\eta^{had}-\eta)^2 + (\varphi^{had}-\varphi)^2 \le R^2.
$$
We took $R=0.4$ and $E^{max}=1$ GeV as in the Tevatron experimental data.
The isolation not only reduces the background from the secondary photons produced by the decays of $\pi^0$ and $\eta$ mesons but also significantly reduces the
so called fragmentation components, connected with collinear photon radiation (10$\%$).

Similarly to the traditional QCD approach the calculated cross section split into two pieces:
$$     
    d\sigma = d\sigma_{direct}(\hat\mu^2) +
d\sigma_{fragm}(\hat\mu^2),
$$
where $d\sigma_{direct}(\hat\mu^2)$ is the perturbative 
contribution, $d\sigma_{fragm}(\hat\mu^2)$ is the
fragmentation contribution, and $\hat\mu^2$ is the fragmentation scale. In our calculations $\hat\mu$ is the invariant mass of the produced photon and any final quark
and we restrict the direct contribution to $\hat\mu > M
= 1$ GeV in order to eliminate the collinear divergences in the direct cross section. Then the mass of light quark can be safely sent to zero. 
 
\section{Numerical results}
In Figs. 1 -- 3 some of the results of our calculation~\cite{LMZ} are shown (more details see in~\cite{LMZ}). We have found that the full set of
experimental data is reasonably well described by the
$k_T$-factorization approach. One can see that the property of the uintegrated parton distribution and the non-vanishing
transverse momentum of the colliding partons lead to a broadering of the photon transverse momentum distributions
in comparision with the collinear pQCD results.
\begin{figure}[t]
  \includegraphics[height=.35\textheight]{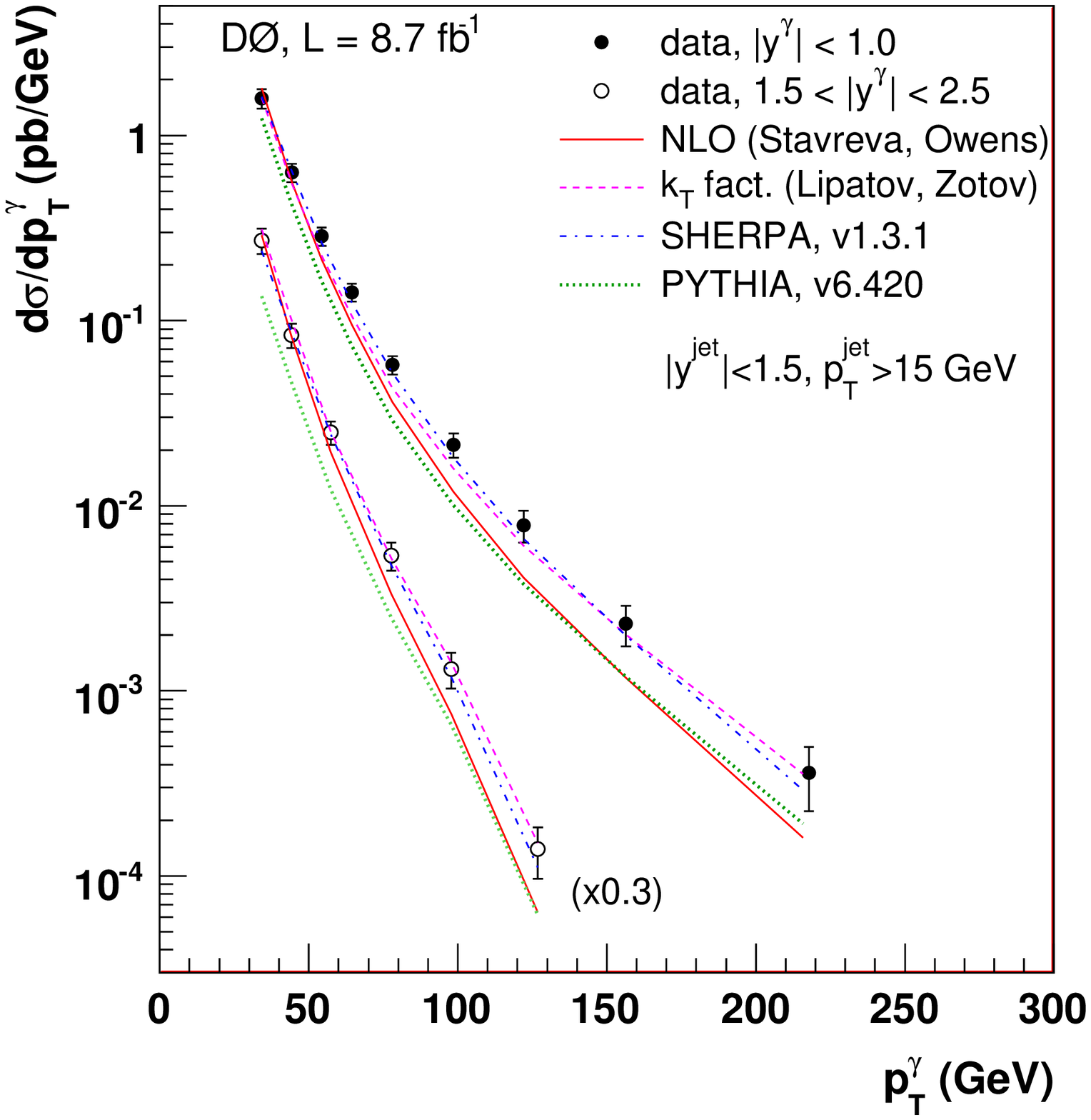}
  \includegraphics[height=.35\textheight]{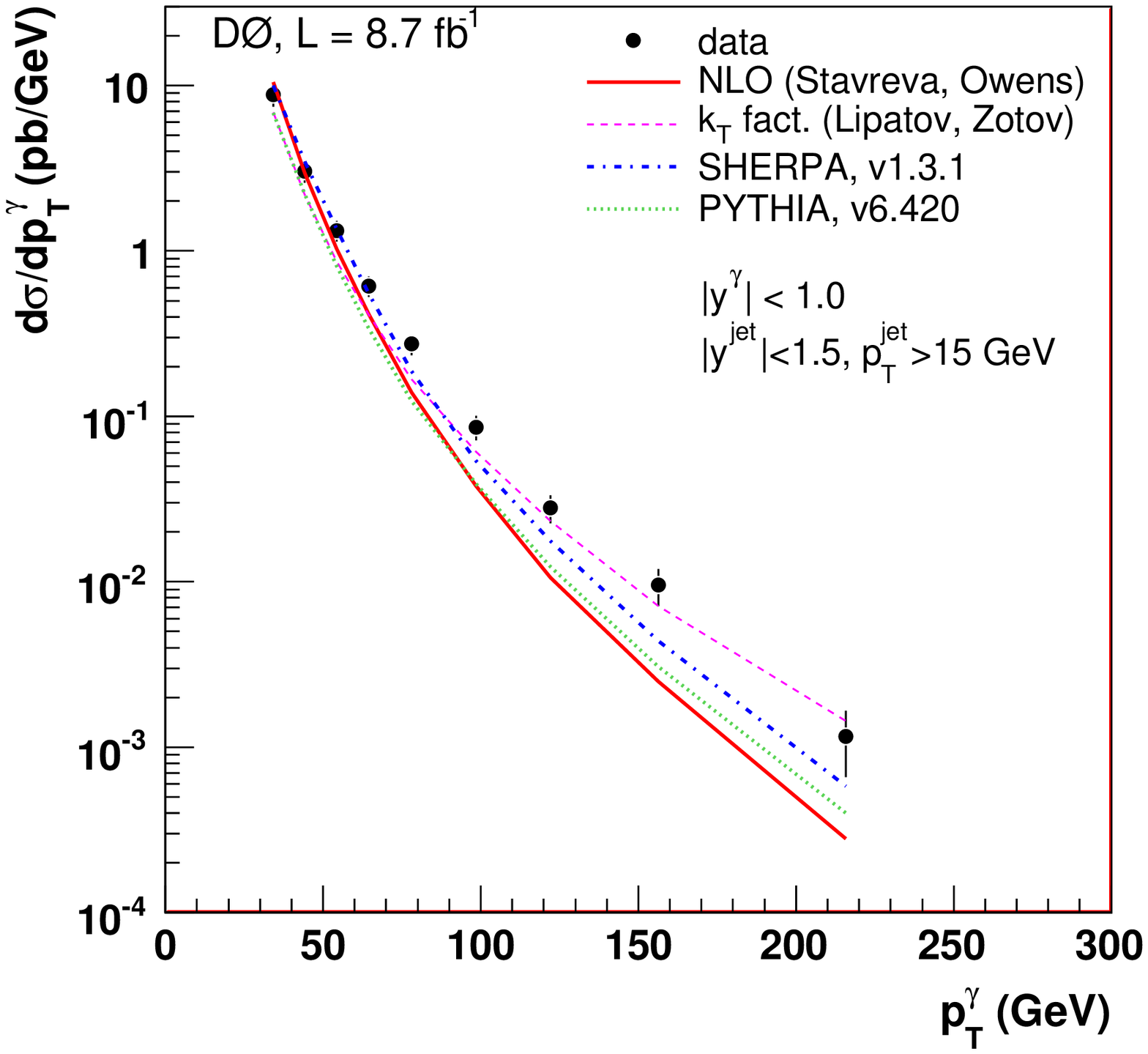}
  \caption{Differential cross section $d\sigma/dp_T^{\gamma}$ of associated $\gamma + b-jet$ (left panel) and $\gamma + c-jet$ (right panel) production at $\sqrt s = 1960$ GeV. Figs.
are taken from~\cite{D0,D0c}.
  }
  \label{fig3}
\end{figure}
As it was noted in~\cite{D0,D0c} our results agree better with the Tevatron data than the NLO QCD ones (see Fig. 3)

The authors would like to thank DESY Directorate for the support in the framework of Moscow
 --- DESY project on MC implementation for HERA --- LHC. A.L. and M.M.
were supported in part by the grant of the president of the Russian Federation (MK-3977.2011.2) and RFBR grant 12-02-31030. This research was supported by the FASI of the Russian Federation (grant NS-3920.2012.2), FASI state contract 02.740.11.0244, RFBR grant 11-02-01454-a, the RMES (grant the Scientific Research on High Energy Physics) and the Ministry of education and sciences of Russia (agreement 8412). N.Z. is very grateful to the Organization Committee, in particular A. Papa and R. Fiore,  for the financial support.




\bibliographystyle{aipproc}   


\end{document}